\documentclass[10pt,a4paper]{article}

\usepackage[utf8]{inputenc}
\usepackage[T1]{fontenc}
\usepackage{lmodern}
\usepackage{amsmath,amssymb}
\usepackage{graphicx}
\usepackage{booktabs}
\usepackage{subcaption}
\usepackage[pagebackref,breaklinks,colorlinks,linkcolor=black,citecolor=black,urlcolor=black]{hyperref}
\usepackage[round]{natbib}
\usepackage{xcolor}
\usepackage{textcomp}
\usepackage[font=small,labelfont=bf]{caption}

\makeatletter
\renewenvironment{abstract}{%
  \small
  \begin{center}%
    \bfseries \abstractname\vspace{-.5em}\vspace{\z@}%
  \end{center}%
  \quotation\noindent\ignorespaces}%
  {\endquotation}
\makeatother

\makeatletter
\renewcommand{\maketitle}{%
  \begin{center}%
    {\LARGE\bfseries \@title \par}%
    \vskip 1.5em%
    {\normalsize \@author \par}%
  \end{center}%
  \par\vskip 1.5em%
}
\makeatother

\title{Rotation-invariant graph message passing enables acquisition protocol generalisation in learning-based brain microstructure estimation}

\author{
  Leevi Kerkel\"a (\href{mailto:leevi.kerkela.17@ucl.ac.uk}{\texttt{leevi.kerkela.17@ucl.ac.uk}}) and Hui Zhang\\[0.5em]
  {\small UCL Hawkes Institute and Department of Computer Science,\\
  University College London, UK}
}

\date{}

\begin{document}
\null\vfill
\maketitle
\thispagestyle{empty}

\begin{abstract}
Estimating brain microstructure has important applications in medicine and neuroscience.
Diffusion-weighted magnetic resonance imaging enables measuring microstructure \textit{in vivo}.
Conventional biophysical model fitting can be accurate but is slow and impractical for time-critical clinical use, where machine learning can offer a potential route to rapid estimation.
We address the problem of microstructure estimation under arbitrary acquisition protocols where most existing learning-based methods fail due to protocol assumptions, requiring retraining when the protocol changes.
We present a graph neural network that represents input data as a point cloud in the 3D space where diffusion-weighted measurements are made and performs rotation-invariant message passing with permutation-invariant pooling, producing fixed-size embeddings that encode microstructure.
The inductive biases of our relatively small model were guided by the underlying physics and symmetries of the problem rather than by generic model architectures.
Trained on randomised simulated data, our model demonstrates domain generalisation, accurately estimating microstructure from data with unseen real-world protocols without retraining.
This approach represents a step towards a \textit{``train once, deploy anywhere''} architecture, bringing rapid learning-based microstructure mapping closer to clinical deployment.
\end{abstract}

\vfill\null

\newpage
\section{Introduction}

Assessing brain tissue microstructure, the organisation of cellular structures at the micrometre scale, is important in neuroscience and medicine. However, conventional biopsy is highly invasive in the brain and therefore rarely feasible, so advances in non-invasive imaging that enable forms of virtual histology are particularly valuable. Diffusion-weighted magnetic resonance imaging (dMRI) is sensitive to thermal-motion-driven displacements of water molecules at the microscopic level, providing a unique way to probe tissue microstructure \textit{in vivo} \citep{johansen2013diffusion}.

Two broad families of methods exist for inferring microstructure from the measured dMRI signals \citep{novikov2019quantifying}. Signal models describe the signal at increasing levels of complexity \citep{basser1994mr,jensen2005diffusional,westin2016q}. In parallel, biophysical models have been proposed to relate the signal more directly to microstructural tissue features such as axon density \citep{szafer1995theoretical}. A notable example is neurite orientation dispersion and density imaging (NODDI) \citep{zhang2012noddi}. Biophysical models provide interpretable microstructural parameters, but fitting them requires computationally expensive and noise-sensitive iterative non-linear optimisation at each voxel. As a result, conventional model fitting can take hours per scan on a standard workstation computer, limiting the clinical feasibility of advanced diffusion modelling when timely results are required.

Recent efforts have focused on accelerating and improving the accuracy of dMRI data processing and microstructure estimation pipelines using machine learning \citep{karimi2024diffusion}. For example, deep learning methods for data processing such as eddy-current distortion correction have achieved order-of-magnitude speedups compared to conventional algorithms \citep{legouhy2024eddeep}. At the parameter-estimation stage, scanner-integrated machine learning has enabled near real-time parameter mapping, for instance, generating NODDI maps during image acquisition in seconds compared to hours \citep{rot2025real}. However, most learning-based microstructure estimation techniques are tied to a specific acquisition protocol, do not take protocol settings as model inputs, and generalise poorly to unseen diffusion-encoding magnitudes or directions without retraining. Models requiring fixed-size inputs cannot be applied to datasets with a different number of acquisitions. Furthermore, standard deep learning network architectures do not enforce the physical symmetries of dMRI: diffusion is antipodally symmetric, scalar microstructural parameters are independent of orientation, and the order of acquisitions does not matter.

In this work, we present a graph neural network (GNN) operating at voxel level that addresses these challenges. Each measurement is treated as a point in the 3D space where dMRI measurements are made, and each point connects to its neighbours to produce a graph. In this representation, each point encodes the acquisition settings of a measurement by its position. The network performs message passing followed by pooling to obtain a fixed-size embedding encoding microstructure. Physical symmetries are enforced by construction rather than learned from augmented data, so our design meets the key requirements for protocol generalisation: first, it accommodates a variable number of inputs, and second, it is invariant by construction to reflections, rotations, and permutations of the input point cloud. Our model achieves state-of-the-art accuracy across protocols unseen during training, demonstrating the feasibility of \textit{``train once, deploy anywhere''} learning-based microstructural imaging (in principle; we do not claim universal deployment of this specific model).

Our key contributions are:
\begin{itemize}
\item To our knowledge, the first learning-based microstructure estimation method that generalises over acquisition protocols without making assumptions about sampling (\textit{e.g.}, spherical shells)
\item A GNN designed for microstructural imaging with the desirable invariance properties by construction
\item Embeddings that smoothly reflect microstructure across protocols despite not having any explicit cross-protocol alignment loss in training
\end{itemize}

\section{Related work}

\subsection{Geometric deep learning}

Geometric deep learning encodes the relevant symmetry group as an inductive bias so that equivariance holds by construction, rather than being learned from augmented data \citep{bronstein2021geometric}.
In practice, this often means layers that transform equivariantly and applying pooling for invariance.
Set-based architectures build permutation invariance into neural networks while accommodating variable-length inputs \citep{qi2017pointnet,zaheer2017deep}.
GNNs offer a general framework for learning on irregular graph data \citep{scarselli2008graph}, with the message passing formulation \citep{gilmer2017neural} being particularly influential.
$\mathrm{E}(3)$-equivariant GNNs ensure equivariance in learned representations \citep{satorras2021n}.
Spherical convolutional neural networks (SCNNs) \citep{cohen2018spherical,esteves2018learning} implement $\mathrm{SO}(3)$-equivariant convolutions on spherical signals.

\subsection{Learning-based microstructure estimation}

Early work showed that supervised learning from subsampled signals can substantially reduce both scan and processing time for microstructure estimation \citep{golkov2016q}.
Subsequent methods sought to mitigate protocol dependence, including projecting normalised $q$-space signals onto a grid to handle varying acquisition protocols under a fixed number of shells (\textit{i.e.}, acquisition points on concentric spheres) \citep{park2021diffnet}, using attention to generalise across diffusion-encoding directions \citep{zong2024attention}, and employing a set encoder to enable continuous signal reconstruction at arbitrary $q$-space locations \citep{ewert2024geometric}.
However, these architectures are not inherently rotation invariant.
Convolution and transformer GNNs have been applied to estimate microstructure from subsampled data under a fixed protocol (in contrast to our focus on protocol generalisation) \citep{chen2020estimating,chen2022hybrid}.
Because dMRI measurements are commonly acquired on spherical shells, SCNNs have been applied to improve accuracy, robustness to rotations, and generalisation across diffusion-encoding directions \citep{sedlar2021spherical,sedlar2021diffusion,goodwin2022can,kerkela2024spherical,elaldi2024equivariant}.
However, while SCNNs accommodate varying directions within a shell, existing architectures generally assume $q$-space sampling on fixed shells and thus do not generalise across varying shell counts or to non-spherical sampling schemes.
To the best of our knowledge, no prior work has trained a single model that generalises to arbitrary protocols.

\section{Preliminaries}
\label{sec:preliminaries}

\begin{figure}[t]
  \centering
  \begin{subfigure}{0.32\linewidth}
    \includegraphics[width=\linewidth]{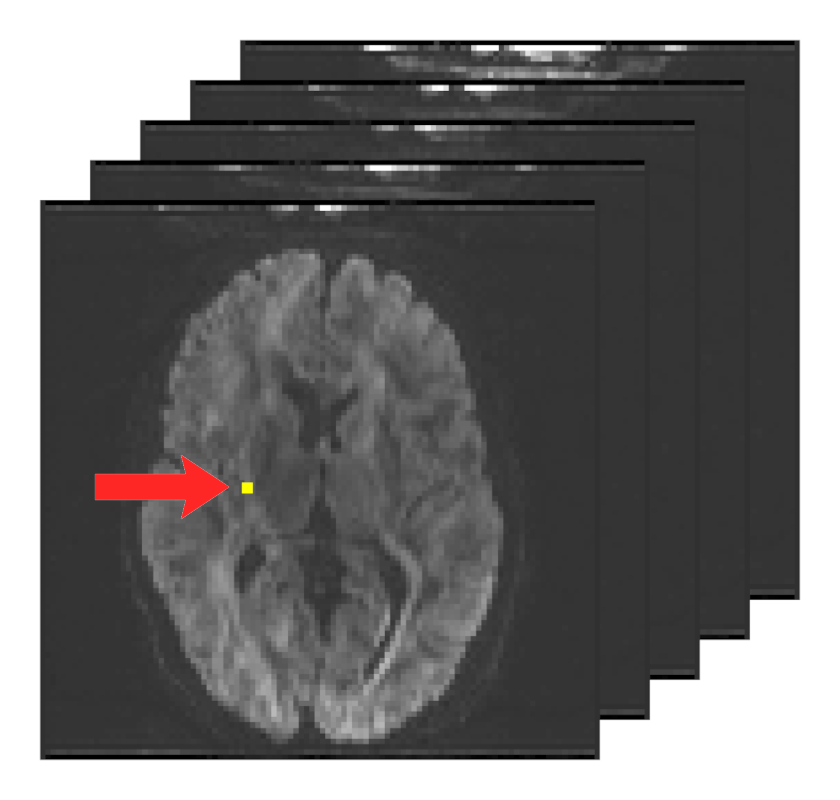}
    \caption{}
    \label{fig:dmri_data}
  \end{subfigure}
  \hfill
  \begin{subfigure}{0.32\linewidth}
    \includegraphics[width=\linewidth]{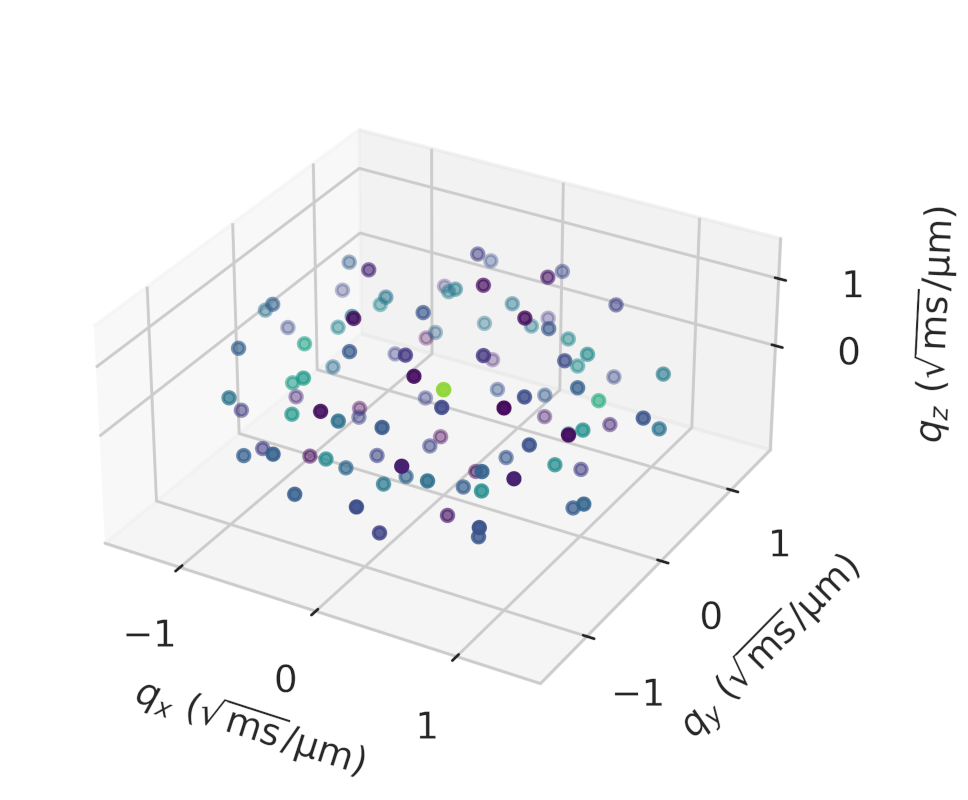}
    \caption{}
    \label{fig:voxel-level_dmri_data}
  \end{subfigure}
  \hfill
  \begin{subfigure}{0.32\linewidth}
    \includegraphics[width=\linewidth]{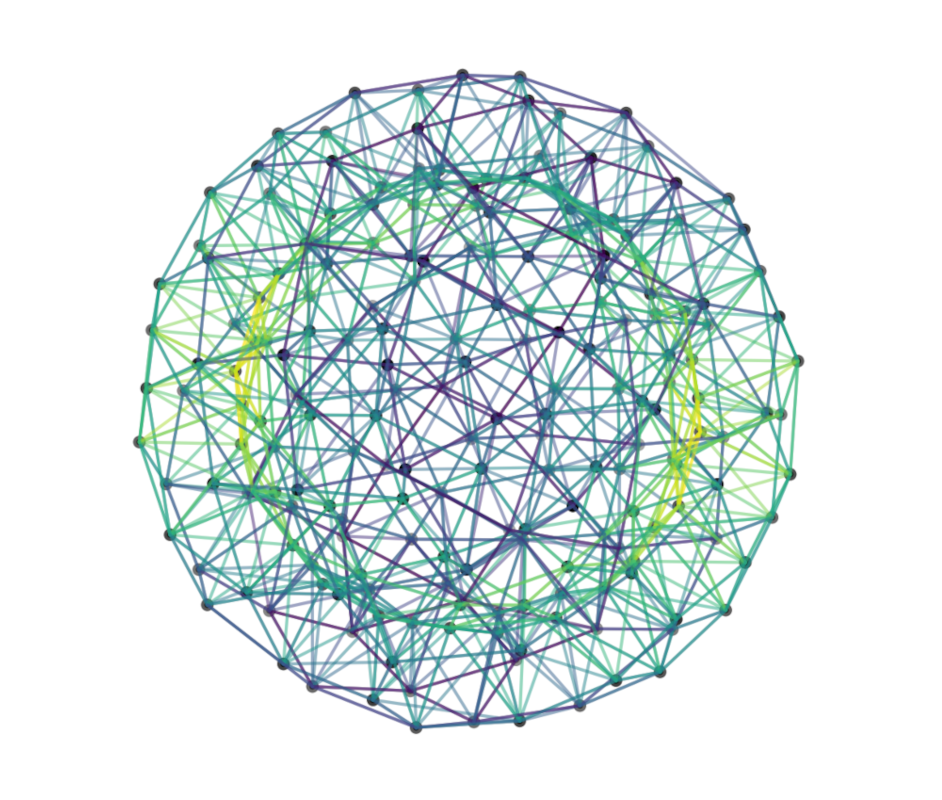}
    \caption{}
    \label{fig:model_input_graph}
  \end{subfigure}
  \caption{
Illustration of the model input data.
(a) Axial slices from five dMRI data volumes, each acquired with specific acquisition settings.
(b) The measurements from the voxel highlighted in (a) across all volumes in the 3D space where the dMRI measurements are made, revealing the geometric structure of the two-shell protocol ($b$ = 1 and 2 ms/\textmu m$^2$, 50 directions per shell); colour encodes relative signal intensity.
(c) The graph constructed from the voxel-level data in (b) used as the GNN input: each measurement is mirrored across the origin to enforce antipodal symmetry, and a $k$-nearest neighbour graph ($k = 8$) is constructed; colour encodes edge feature norm magnitude.
  }
\end{figure}

\subsection{Diffusion MRI data}

Data from a typical dMRI scan (Figure~\ref{fig:dmri_data}) consists of a series of 3D volumes, each acquired with specific acquisition settings, such that the diffusion encoding in volume $i$ is characterised by the $b$-value $b_i \in \mathbb{R}_+$ and diffusion-encoding direction $\mathbf{g}_i \in S^2$.
The measured signal $S_i \in \mathbb{R}_+$ depends on both.

It is convenient to describe acquisitions in terms of $q$-space, the Fourier dual of the water displacement distribution \citep{stejskal1965spin}.
Ignoring physical constants and the effects of time-dependent diffusion, and using units of ms/\textmu m$^2$ for the $b$-value, we have $\mathbf{q}_i = \sqrt{b_i}\, \mathbf{g}_i \in \mathbb{R}^3$.
This scaling yields $q$-space coordinates of order unity for typical experiments.
Because diffusion is antipodally symmetric, $\mathbf{q}_i$ and $-\mathbf{q}_i$ correspond to the same diffusion encoding.
We define the protocol as
\begin{equation}
\mathcal{Q} = \{ \mathbf{q}_i \}_{i=1}^N ,
\end{equation}
where $N$ is the number of acquisitions, which varies across protocols.

At the voxel level, data can thus be viewed as a point cloud (Figure~\ref{fig:voxel-level_dmri_data})
\begin{equation}
\label{eq:dmri_data}
\mathcal{D} = \{ (\mathbf{q}_i, S_i) \}_{i=1}^N,
\end{equation}
where the ordering is irrelevant.
This motivates learning formulations that treat dMRI data as a set rather than as a fixed-length ordered vector.

\subsection{Problem statement}

The goal of microstructural imaging is to estimate clinically and scientifically relevant microstructural properties. A biophysical forward model $f$ expresses the signal as
\begin{equation}
\hat{S}_i = f(\mathbf{q}_i, \boldsymbol{\theta}),
\end{equation}
where $\hat{S}_i \in \mathbb{R}_+$ is the predicted signal and $\boldsymbol{\theta} \in \Theta \subset \mathbb{R}^d$ are microstructural parameters with $d$ being the number of parameters, and $\Theta$ is the space of biologically plausible values.

Conventional fitting usually solves
\begin{equation}
\hat{\boldsymbol{\theta}} = \operatorname*{arg\,min}_{\boldsymbol{\theta} \in \Theta} \sum_{i=1}^N \left( S_i - \hat{S}_i \right)^2 .
\end{equation}
Other formulations, such as maximum-likelihood estimators that explicitly model Rician noise \citep{landman2007diffusion}, also exist but are conceptually similar. In the learning setting, we train a direct mapping
\begin{equation}
\hat{\boldsymbol{\theta}} = F_\phi(\mathcal{D}),
\end{equation}
where $\phi$ are learnable parameters. Here, our goal is to train $F_\phi$ that satisfies
\begin{equation}
F_\phi\!\big(\{ ( Q \mathbf{q}_{\pi(i)}, S_{\pi(i)} ) \}_{i=1}^N\big)
=
F_\phi\!\big(\{ ( \mathbf{q}_i, S_i ) \}_{i=1}^N\big)
\end{equation}
for any orthogonal transformation $Q \in \mathrm{O}(3)$ (including rotations and reflections) and any permutation $\pi \colon \{1,\ldots,N\} \to \{1,\ldots,N\}$, and that can be applied to data acquired with various protocols to obtain accurate $\hat{\boldsymbol{\theta}}$.

\section{Method}

Our GNN comprises graph construction, message passing layers propagating information between points, pooling to obtain a fixed-size embedding, and microstructural parameter prediction via a multi-layer perceptron (MLP). To guarantee invariance to orthogonal transformations and permutations of $\mathcal{D}$, we restrict the node and edge features to functions of signals, $b$-values, pairwise Euclidean distances, and angular differences.

\subsection{Graph construction}

We first normalise the signals by measurements with $b{=}0$ (a standard practice because diffusion is measured from the relative differences between signals with different $b$), yielding $E_i = S_i \big/ \left( \frac{1}{|\mathcal{B}_0|} \sum_{j \in \mathcal{B}_0} S_j \right)$, where $\mathcal{B}_0 = \{ j \mid b_j = 0 \}$.

We keep the points with non-zero $b$ as nodes in the graph and exploit the antipodal symmetry of diffusion: for each $(\mathbf{q}_i, E_i)$, we add $(-\mathbf{q}_i, E_i)$, creating $M = 2(N - |\mathcal{B}_0|)$ nodes in total. Each node has initial features $\mathbf{x}_i^{(0)} = [E_i;\, b_i]$, where the superscript denotes the layer index. Signals and $b$-values are invariant to orthogonal transformations because they are independent of the $q$-space coordinates.

Edges are defined via $k$-nearest neighbours in $q$-space, producing an undirected graph (Figure~\ref{fig:model_input_graph}). For each node $i$, we denote the set of its neighbours by $\mathcal{N}(i)$ which is invariant to orthogonal transformations and permutations of $\mathcal{D}$ because it depends only on pairwise distances, which are preserved under orthogonal transformations:
$\lVert Q\mathbf{q}_i - Q\mathbf{q}_j \rVert = \sqrt{(\mathbf{q}_i - \mathbf{q}_j)^\top Q^\top Q (\mathbf{q}_i - \mathbf{q}_j)} = \lVert \mathbf{q}_i - \mathbf{q}_j \rVert$,
since $Q^\top Q = I$ for all $Q \in \mathrm{O}(3)$.

Edge features are then defined for each $i$ and $j \in \mathcal{N}(i)$ as
\begin{equation}
\mathbf{e}_{ij} = \big[\lVert \mathbf{q}_i - \mathbf{q}_j \rVert ; \left| \frac{\mathbf{q}_i \cdot \mathbf{q}_j}{\lVert \mathbf{q}_i \rVert \lVert \mathbf{q}_j \rVert} \right| ; |b_i - b_j|\big],
\end{equation}
where $[\,;\,]$ denotes concatenation and $\lVert \cdot \rVert$ the Euclidean norm, and the absolute value of the cosine is because of antipodal symmetry. The edge features are invariant to orthogonal transformations: the pairwise distance and the $b$-value difference depend only on quantities already shown to be invariant, and the cosine similarity satisfies
\begin{equation}
\frac{(Q\mathbf{q}_i)^\top (Q\mathbf{q}_j)}{\lVert Q\mathbf{q}_i \rVert \lVert Q\mathbf{q}_j \rVert}
= \frac{\mathbf{q}_i^\top Q^\top Q\, \mathbf{q}_j}{\lVert \mathbf{q}_i \rVert \lVert \mathbf{q}_j \rVert}
= \frac{\mathbf{q}_i^\top \mathbf{q}_j}{\lVert \mathbf{q}_i \rVert \lVert \mathbf{q}_j \rVert} .
\end{equation}

\subsection{Message passing}

We use permutation-invariant aggregation and update node features using only the invariant node and edge features defined above. 

The first message passing layer takes node and edge features:
\begin{equation}
\mathbf{m}_{ij}^{(1)} = \mathrm{MLP}_{\text{edge}}^{(1)}\left([\mathbf{x}_j^{(0)}; \mathbf{e}_{ij}]\right).
\end{equation}
The subsequent layers operate only on updated node features given by the previous layer since edge geometry is already encoded in the first layer:
\begin{equation}
\mathbf{m}_{ij}^{(\ell)} = \mathrm{MLP}_{\text{edge}}^{(\ell)}\!\left(\mathbf{x}_j^{(\ell-1)}\right), \quad \ell \ge 2,
\end{equation}
where $\ell$ denotes the layer index.

At each layer, messages are mean-aggregated
\begin{equation}
\bar{\mathbf{m}}_{i}^{(\ell)} = \frac{1}{|\mathcal{N}(i)|} \sum_{j \in \mathcal{N}(i)} \mathbf{m}_{ij}^{(\ell)},
\end{equation}
and these are concatenated to node features and fed to a node MLP to obtain the updated node features:
\begin{equation}
\mathbf{x}_i^{(\ell)}=\mathrm{MLP}_{\text{node}}^{(\ell)}\!\left([\mathbf{x}_i^{(\ell-1)};\, \bar{\mathbf{m}}_{i}^{(\ell)}]\right).
\end{equation}

\subsection{Pooling and parameter prediction}

To obtain an embedding whose size is independent of the number of nodes, we apply attention-weighted \citep{bahdanau2014neural} pooling after all message passing layers:
\begin{equation}
\mathbf{z} = \sum_{i=1}^{M} w_i \, \mathbf{x}_i^{(L)}, \quad
w_i = \frac{\exp\!\big(\mathrm{MLP}_{\text{attn}}(\mathbf{x}_i^{(L)})\big)}{\sum_{k=1}^{M}\exp\!\big(\mathrm{MLP}_{\text{attn}}(\mathbf{x}_k^{(L)})\big)},
\end{equation}
where $\mathrm{MLP}_{\text{attn}}$ produces scalar attention scores and $L$ is the total number of message passing layers. The final network output is
\begin{equation}
\hat{\boldsymbol{\theta}} = \mathrm{MLP}_{\text{readout}}(\mathbf{z}) .
\end{equation}
Permutation invariance follows from the use of mean aggregation over neighbours and attention-weighted summation over all nodes, both of which are symmetric functions of their inputs \citep{zaheer2017deep}. Since each component---node features, edge features, neighbourhood construction, message passing, and pooling---maps $\mathrm{O}(3)$- and permutation-invariant inputs to invariant outputs, their composition $F_\phi$ is invariant.
\section{Experiments}

We evaluated whether our GNN can accurately estimate NODDI parameters from simulated data with a known ground truth using real-world protocols not seen during training. The NODDI parameters are the neurite density index ($\mathrm{NDI}$), orientation dispersion index ($\mathrm{ODI}$), and free water fraction ($\mathrm{FWF}$).

\subsection{Network architecture}

We used $k=8$ and three message passing layers. Each edge MLP had architecture $d_\text{in} \!\to\! 64 \!\to\! 64 \!\to\! 16$, where $d_\text{in}$ is the number of input features, and each node MLP had $(d_\text{in} + 16) \!\to\! 64 \!\to\! 64 \!\to\! 16$. The attention MLP had $16 \!\to\! 16 \!\to\! 1$. The readout MLP had $16 \!\to\! 32 \!\to\! 3$, intentionally small to promote encoding microstructure in $\mathbf{z}$ instead of a higher-capacity decoder learning complex disentangling. The model had 40,132 learnable parameters. Sigmoid linear unit (SiLU) activations were used throughout.

\subsection{Training data generation}

We simulated 10{,}000 voxels using the NODDI biophysical model. For each voxel, we sampled NDI, ODI, and FWF independently from $\mathcal{U}(0.025, 0.975)$, and sampled the principal fibre direction from $\mathcal{U}(S^2)$. For protocol generalisation, we generated 10 random protocols per voxel, yielding 100{,}000 training examples. For each protocol, the number of shells was uniformly sampled from $\{2,3,4,5\}$, the $b$-values from $\mathcal{U}(0.25, 5)$ ms/\textmu m$^2$, and the number of gradient directions between 12 and 128 independently per shell, with directions drawn uniformly over the hemisphere. We added Rician noise with SNR sampled per batch from $\mathcal{U}(10, 40)$, so the model never saw the same data twice.

\subsection{Training}

The model was trained for 500 epochs with Adam (initial learning rate $10^{-3}$, multiplied by 0.99 every fifth epoch), batch size 10, and gradient clipping with a maximum norm of 1. For protocol generalisation, data were organised so that each group of 10 successive batches had the same microstructural parameters but different protocols. We accumulated gradients over these 10 batches before an optimiser step. For each voxel, we randomly dropped a fraction of measurements sampled from $\mathcal{U}(0, 0.5)$ with probability 0.5 to avoid training only on data with complete spherical sampling. The loss was mean squared error (MSE). The model was implemented using PyTorch Geometric \citep{fey2019fast}.

\subsection{Evaluation}

We evaluated on three real-world protocols that differ substantially, providing a realistic test of cross-protocol generalisation. We simulated data using a Diffusion Spectrum Imaging (DSI) protocol with 303 points on a Cartesian grid (\textit{i.e.}, not a protocol with shells) covering a hemisphere with maximum $b$-value of 5 ms/\textmu m$^2$ \citep{radhakrishnan2024practical}, the Human Connectome Project (HCP) 3-shell protocol ($b = 1, 2, 3$ ms/\textmu m$^2$, 90 directions per shell) \citep{sotiropoulos2013advances}, and the UK Biobank (UKBB) 2-shell protocol ($b = 1, 2$ ms/\textmu m$^2$, 50 directions per shell) \citep{alfaro2018image}. SNR was 30 for DSI, 25 for HCP, and 20 for UKBB. We evaluated accuracy on a \(5\times5\times5\) grid of parameter values $(\mathrm{NDI}, \mathrm{ODI}, \mathrm{FWF}) \in \{0.1, 0.3, 0.5, 0.7, 0.9\}^3$.

\subsection{Baselines}

We compared against two baselines. First, the MATLAB NODDI toolbox\footnote{\url{http://mig.cs.ucl.ac.uk/index.php?n=Tutorial.NODDImatlab}}, which uses iterative non-linear optimisation and is considered the state-of-the-art method, widely used in neuroscience. Second, we used PointNet \citep{qi2017pointnet} as a generic learning-based baseline with a learnable parameter count of 3.5M (nearly two orders of magnitude more than our GNN), treating each measurement as an independent point with features $[q_{x,i};\, q_{y,i};\, q_{z,i};\, E_i;\, b_i]$. We added points mirrored through the origin to the inputs and used the input/feature T-Nets initialised to identity, but omitted batch normalisation to reduce variance from protocol heterogeneity and the orthogonality regularisation term. The NODDI toolbox serves as the primary reference for accuracy, while PointNet demonstrates the benefit of physics-informed inductive biases over a generic point-set architecture.

\subsection{Results}

\subsubsection{Parameter estimation}\label{sec:accuracy}

\begin{figure}[htbp]
  \centering
  \begin{subfigure}{\linewidth}
    \includegraphics[width=\linewidth]{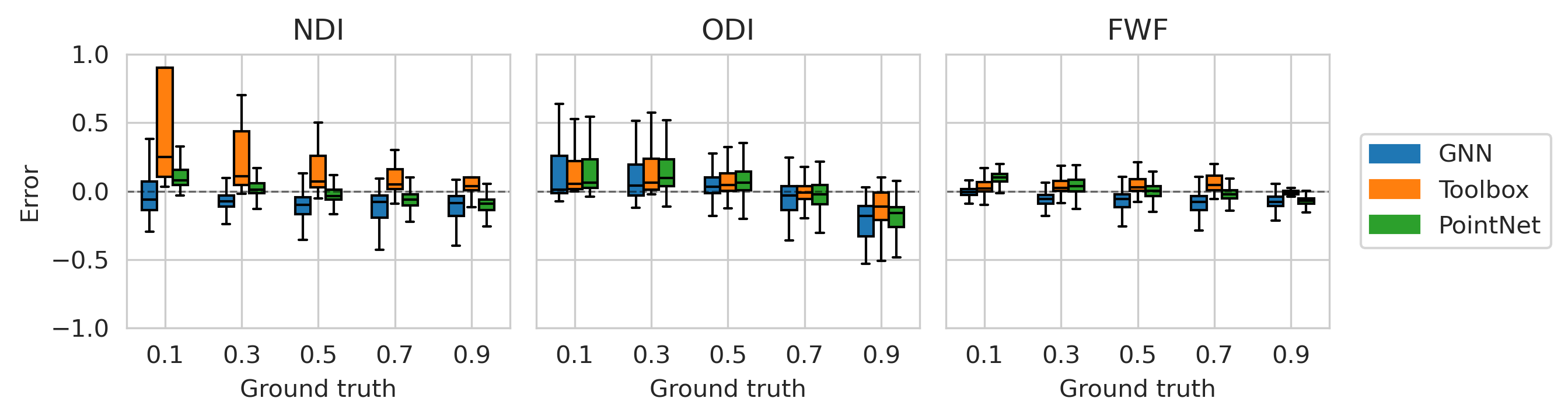}
    \caption{}
    \label{fig:errors_dsi}
  \end{subfigure}

  \medskip
  \begin{subfigure}{\linewidth}
    \includegraphics[width=\linewidth]{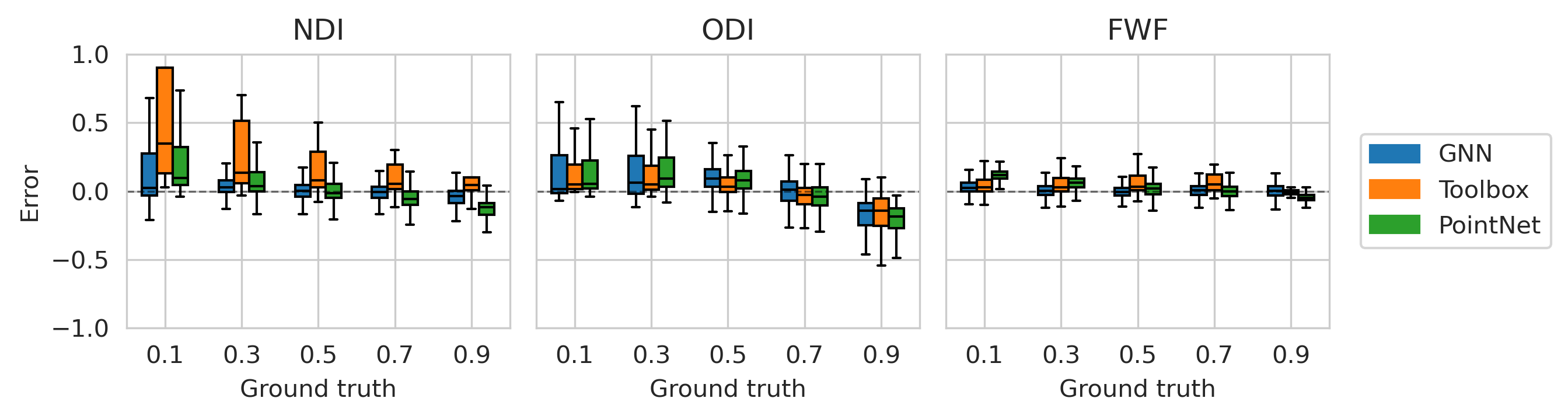}
    \caption{}
    \label{fig:errors_hcp}
  \end{subfigure}

  \medskip
  \begin{subfigure}{\linewidth}
    \includegraphics[width=\linewidth]{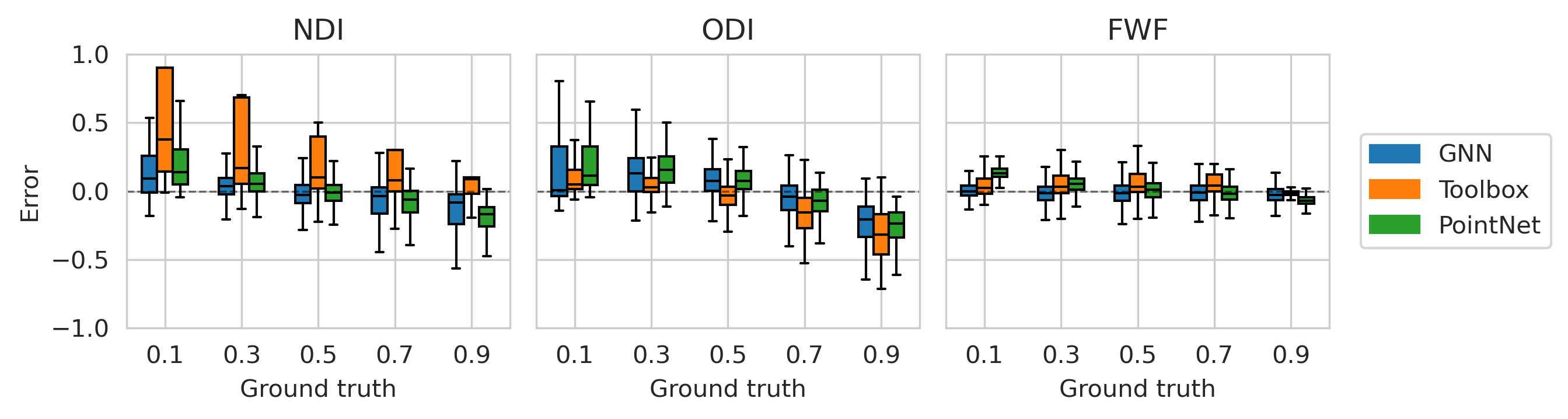}
    \caption{}
    \label{fig:errors_ukbb}
  \end{subfigure}
  \caption{NODDI parameter estimation errors for DSI (a), HCP (b), and UKBB (c) protocols on the test dataset.}
\end{figure}

\begin{table}[htbp]
  \centering
  \scalebox{0.75}{\begin{minipage}{\linewidth}

  \begin{subtable}{\linewidth}
    \centering
    \caption{Across all parameters}
    \begin{tabular}{lccc}
      \toprule
      Protocol & GNN & Toolbox & PointNet \\
      \midrule
      DSI  & 0.024 & 0.048 & \textbf{0.019} \\
      HCP  & \textbf{0.017} & 0.051 & 0.023 \\
      UKBB & \textbf{0.025} & 0.061 & 0.029 \\
      \bottomrule
    \end{tabular}
  \end{subtable}

  \bigskip
  \begin{subtable}{\linewidth}
    \centering
    \caption{\(\mathrm{NDI}\)}
    \begin{tabular}{lccc}
      \toprule
      Protocol & GNN & Toolbox & PointNet \\
      \midrule
      DSI  & 0.035 & 0.111 & \textbf{0.018} \\
      HCP  & \textbf{0.017} & 0.119 & 0.029 \\
      UKBB & \textbf{0.028} & 0.128 & 0.034 \\
      \bottomrule
    \end{tabular}
  \end{subtable}

  \bigskip
  \begin{subtable}{\linewidth}
    \centering
    \caption{$\mathrm{ODI}$}
    \begin{tabular}{lccc}
      \toprule
      Protocol & GNN & Toolbox & PointNet \\
      \midrule
      DSI  & 0.030 & \textbf{0.029} & 0.033 \\
      HCP  & 0.032 & \textbf{0.026} & 0.032 \\
      UKBB & \textbf{0.040} & 0.045 & 0.042 \\
      \bottomrule
    \end{tabular}
  \end{subtable}

  \bigskip
  \begin{subtable}{\linewidth}
    \centering
    \caption{\(\mathrm{FWF}\)}
    \begin{tabular}{lccc}
      \toprule
      Protocol & GNN & Toolbox & PointNet \\
      \midrule
      DSI  & 0.008 & \textbf{0.006} & \textbf{0.006} \\
      HCP  & \textbf{0.003} & 0.009 & 0.007 \\
      UKBB & \textbf{0.006} & 0.010 & 0.010 \\
      \bottomrule
    \end{tabular}
  \end{subtable}
  \end{minipage}}

  \caption{Mean squared errors over the test dataset. The best in bold.}
  \label{tab:MSEs}
\end{table}

For each grid point and protocol, we generated \(100\) independent noise realisations, yielding \(12{,}500\) test voxels per protocol. In terms of total MSE, both learning-based methods outperformed conventional fitting. Table~\ref{tab:MSEs} summarises MSEs for different parameters over the test dataset: the GNN achieved the lowest total loss for HCP and UKBB protocols, while PointNet had lower loss with the DSI protocol. Figures~\ref{fig:errors_dsi}--\ref{fig:errors_ukbb} show estimation error distributions across the parameter space.

\subsubsection{Rotation variance}

To compare the GNN to PointNet in terms of rotation variance, we used the UKBB protocol, set SNR to infinity to exclude the effects of noise, and applied 100 random $\mathrm{SO}(3)$-rotations to the underlying microstructure at each test dataset grid point. We only tested $\mathrm{SO}(3)$ rotations because invariance under reflections is guaranteed by the antipodal mirroring in the graph construction. The GNN with inductive biases exhibited substantially lower rotation variance (mean standard deviation of model outputs over rotated microstructure = 0.004) than PointNet (mean standard deviation = 0.022) that had to learn robustness to rotations. We did not include the toolbox in this comparison as conventional fitting can nearly perfectly invert the NODDI biophysical forward model on noise-free data.

\subsubsection{Embeddings}

\begin{figure}[htbp]
  \centering
  \begin{subfigure}{0.33\linewidth}
    \includegraphics[width=\linewidth]{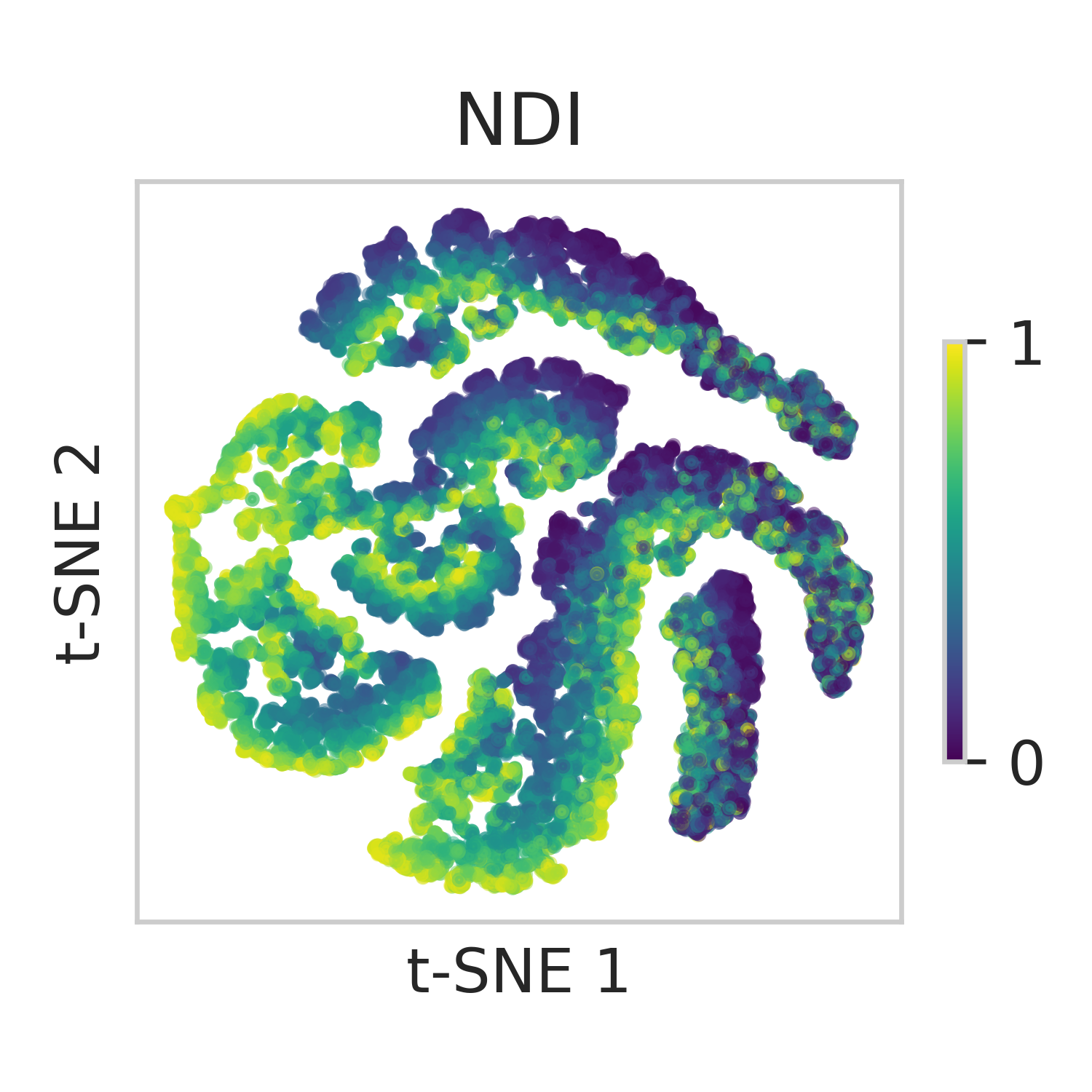}
    \caption{}
    \label{fig:embedding_tsne_1}
  \end{subfigure}
  \begin{subfigure}{0.33\linewidth}
    \includegraphics[width=\linewidth]{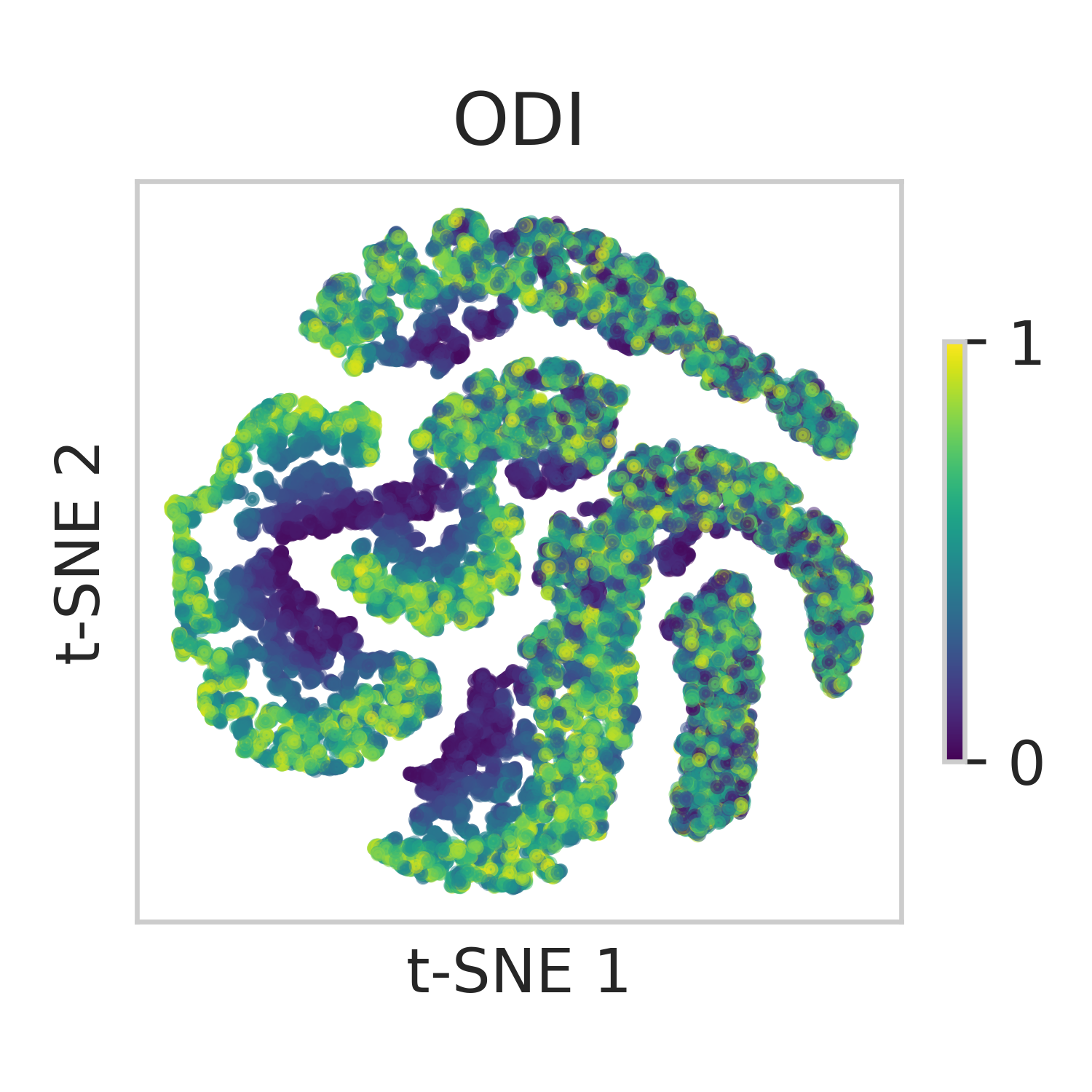}
    \caption{}
    \label{fig:embedding_tsne_2}
  \end{subfigure}

  \medskip
  \begin{subfigure}{0.33\linewidth}
    \includegraphics[width=\linewidth]{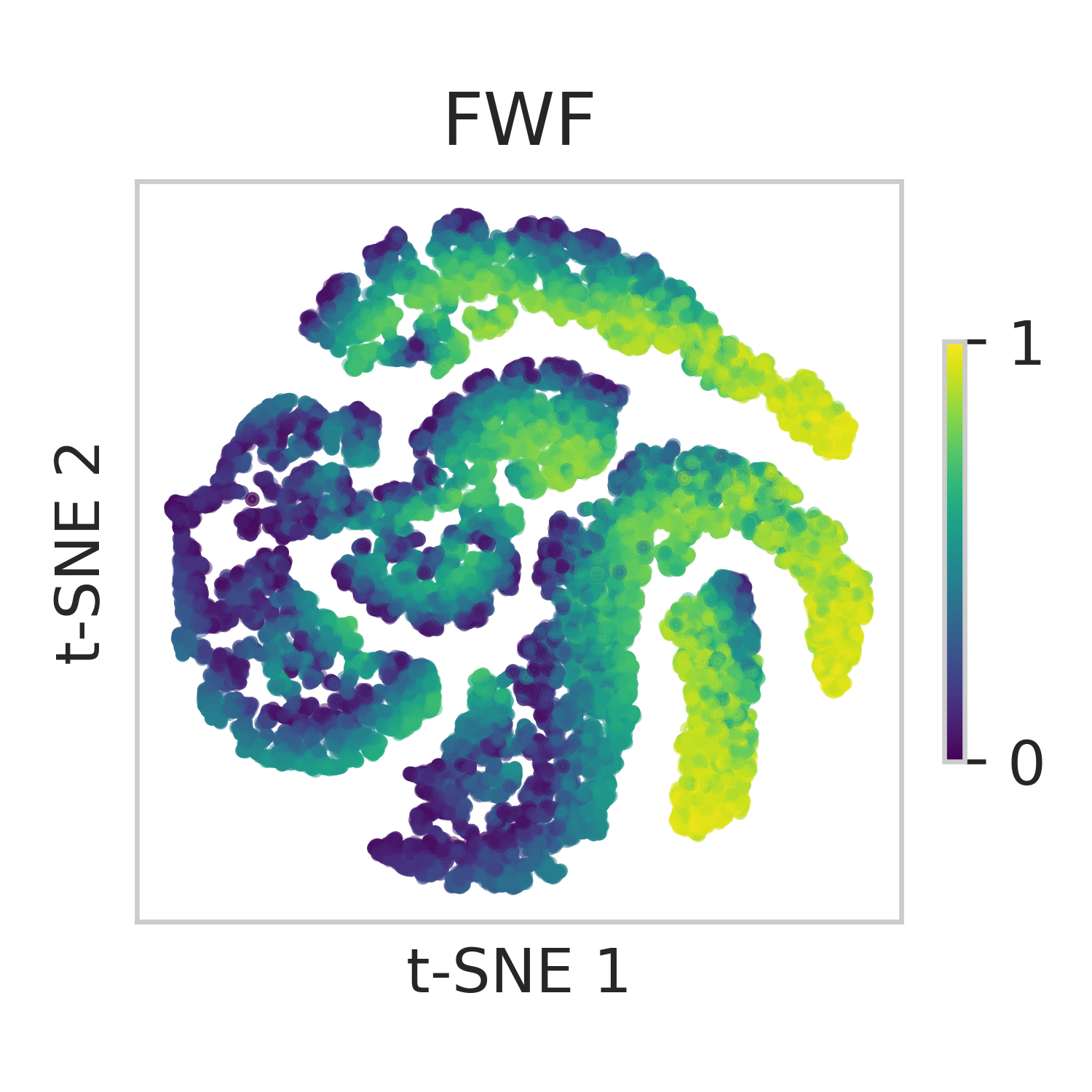}
    \caption{}
    \label{fig:embedding_tsne_3}
  \end{subfigure}
  \begin{subfigure}{0.33\linewidth}
    \includegraphics[width=\linewidth]{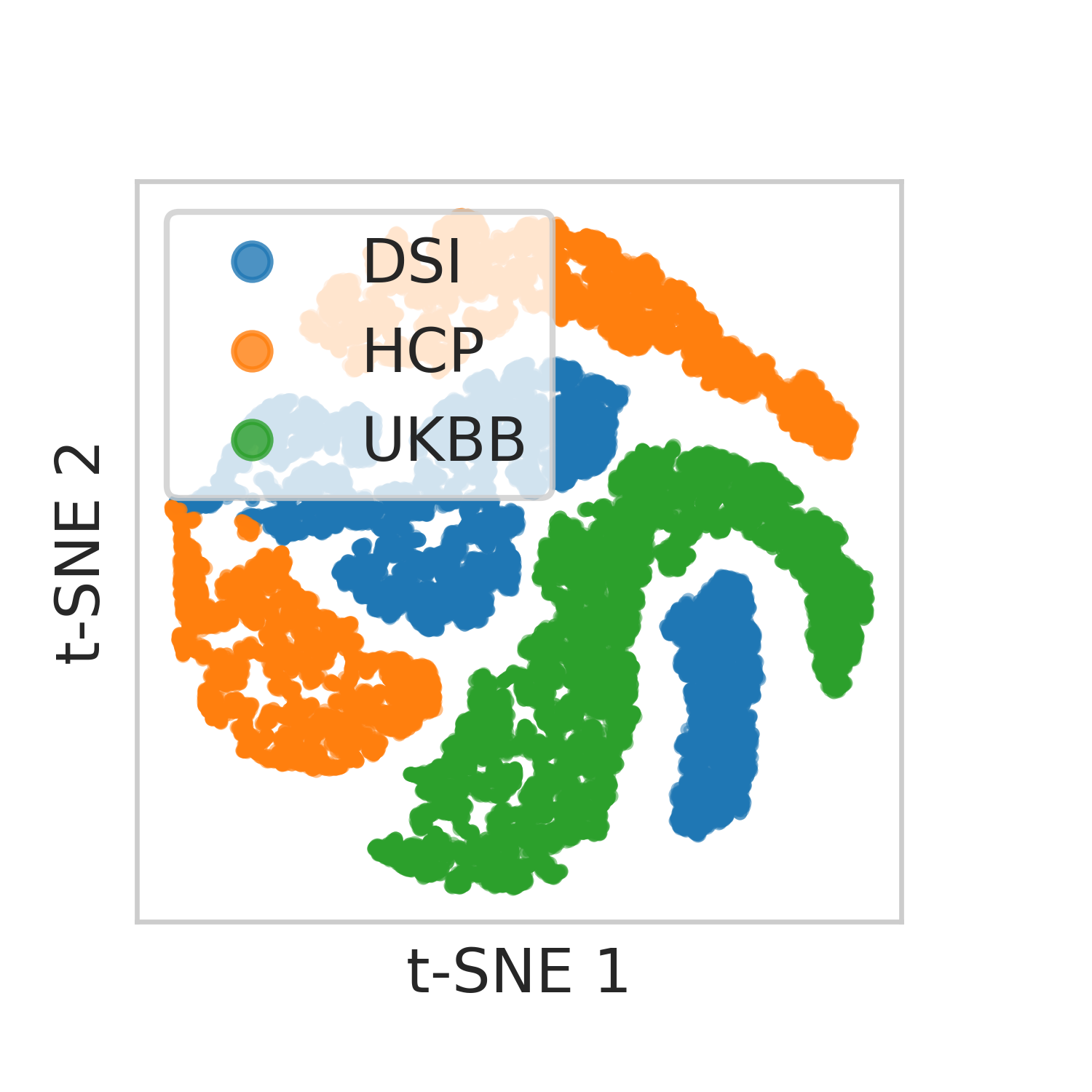}
    \caption{}
    \label{fig:embedding_tsne_protocols}
  \end{subfigure}
\caption{
  t-SNE on the pooled embeddings produced by the GNN from test data with DSI, HCP, and UKBB protocols, revealing microstructure-aligned manifolds.
  (a)--(c) Embeddings coloured by ground-truth parameter values.
  (d) Embeddings coloured by protocol.
}
\end{figure}

To assess the embeddings, we created 10,000 test voxels for each of the three protocols by randomly sampling from the microstructural parameter space like when generating the training data. Principal component analysis (PCA) revealed low-dimensional structure with the first six components explaining 90\% of the variance. The embeddings are visualised in Figures~\ref{fig:embedding_tsne_1}--\ref{fig:embedding_tsne_3} using t-distributed stochastic neighbour embedding (t-SNE), highlighting smooth microstructural variation in the embeddings across varying protocols, especially with respect to $\mathrm{NDI}$ and $\mathrm{FWF}$.

\subsubsection{Application to real imaging data}

\begin{figure}[htbp]
\centering
\begin{subfigure}{.66\linewidth}
\includegraphics[width=\linewidth]{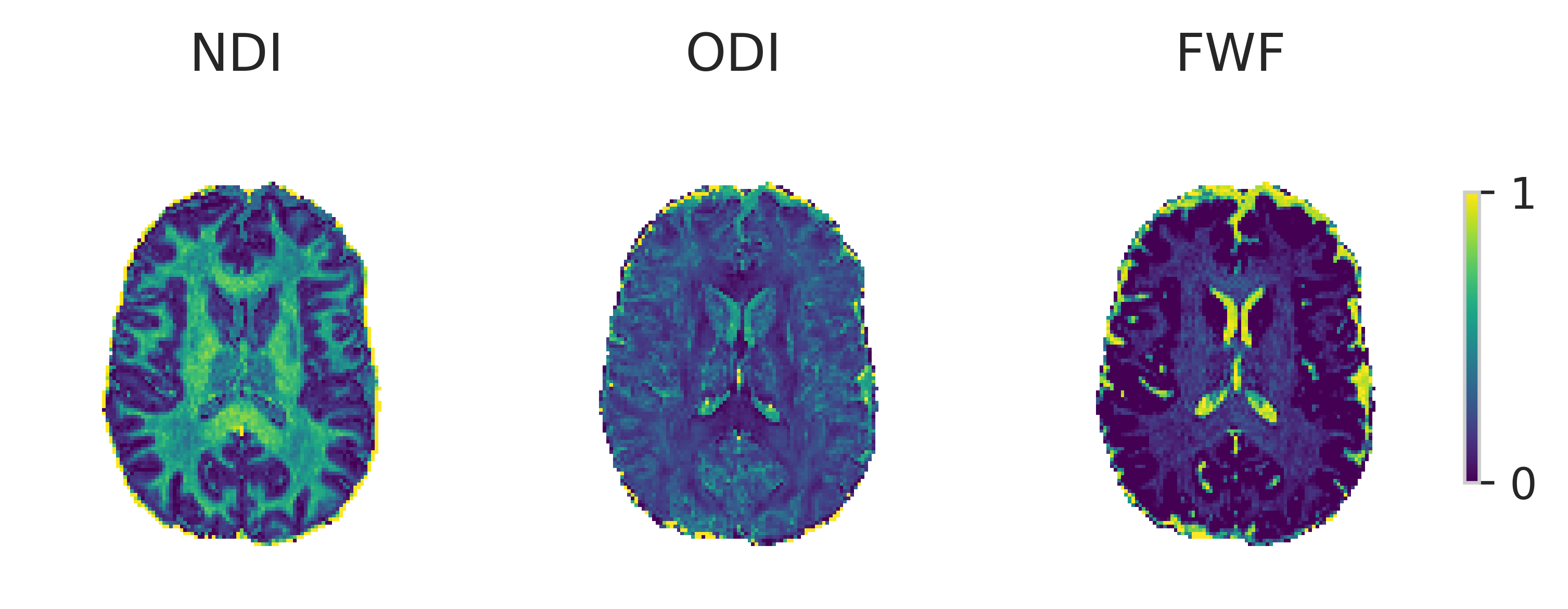}
\caption{}
\label{fig:dsi_maps}
\end{subfigure}

\medskip
\begin{subfigure}{.66\linewidth}
\includegraphics[width=\linewidth]{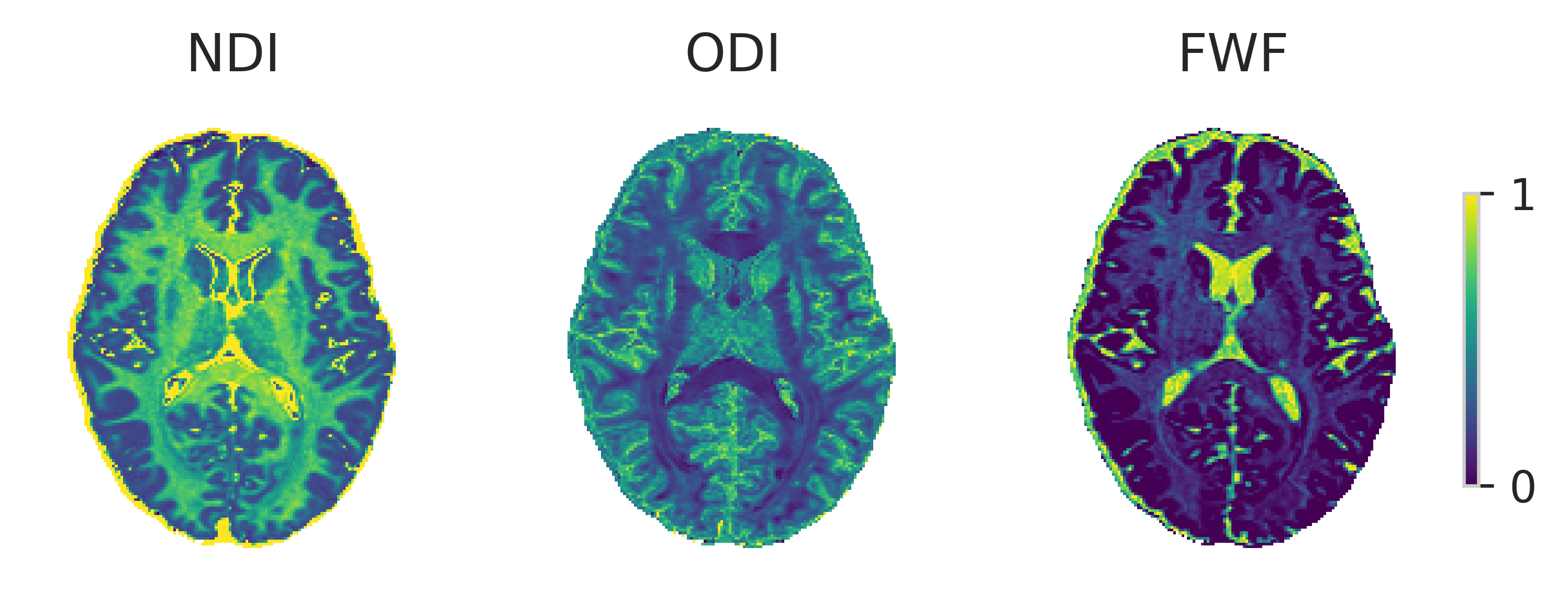}
\caption{}
\label{fig:hcp_maps}
\end{subfigure}

\medskip
\begin{subfigure}{0.66\linewidth}
\includegraphics[width=\linewidth]{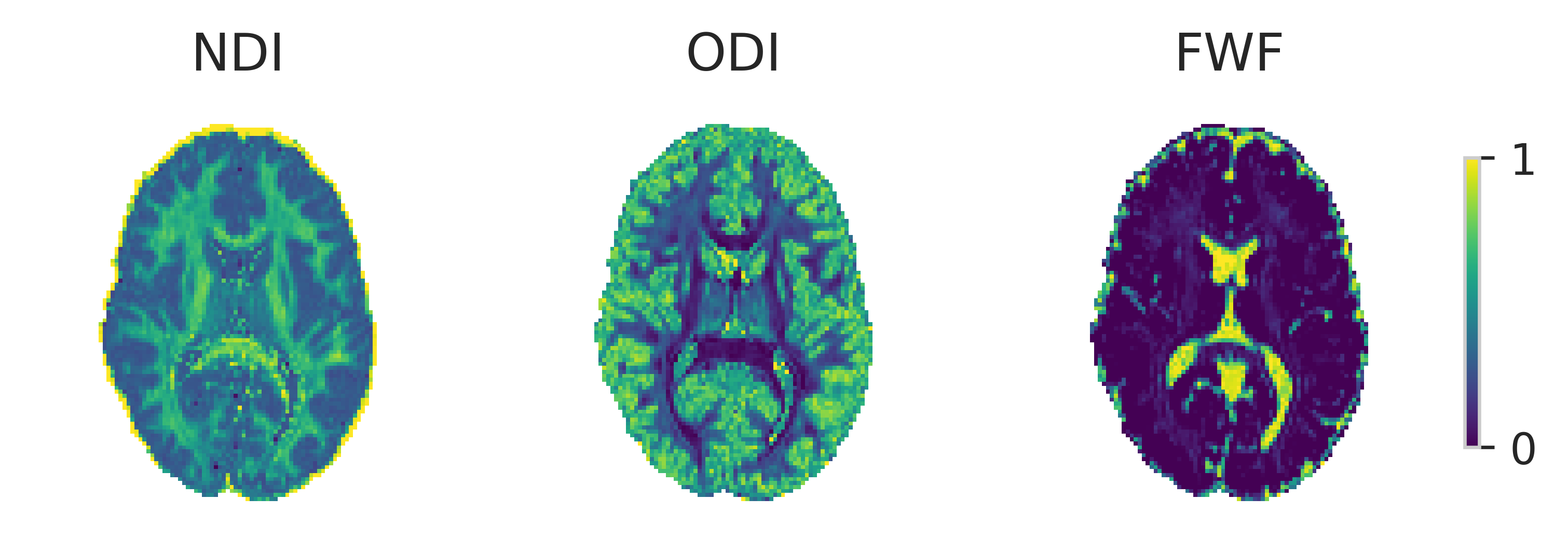}
\caption{}
\label{fig:ukbb_maps}
\end{subfigure}
\caption{Axial slices of NODDI parameter maps produced by the GNN on DSI (a), HCP (b), and UKBB (c) protocols.}
\label{fig:maps}
\end{figure}

Figure~\ref{fig:maps} shows NODDI maps produced by the GNN on real dMRI data with the three protocols. The visible contrast differences across protocols are expected due to limitations of the forward model, as discussed in Section \ref{sec:conclusion}. These results are presented for qualitative illustration; the GNN's ability to accurately invert the forward model was demonstrated in Section~\ref{sec:accuracy}. On a Dell Precision 5680 laptop workstation (Intel i7-13700H, NVIDIA RTX A1000 6\,GB), GNN inference took 0.12 ms per voxel on the UKBB protocol, while the NODDI toolbox required 164 ms per voxel.

\section{Conclusion}
\label{sec:conclusion}

We have introduced a rotation-invariant GNN for microstructural imaging and shown that it can learn embeddings that enable accurate microstructural parameter estimation from data acquired with protocols unseen during training, while being orders of magnitude faster than conventional fitting. To the best of our knowledge, no existing learning-based method addresses protocol-agnostic microstructure estimation, and the GNN's physics-informed inductive biases yield lower errors and substantially lower rotation variance than a larger generic point-set architecture. We emphasise that the proposed framework is general and independent of the specific biophysical forward model; NODDI serves here as an illustrative example.

A limitation is that we used simulated training data. \textit{In vivo}, the true microstructure is unknown and biophysical models only approximate it; at higher $b$-values, real dMRI signals deviate from the NODDI forward model, reflecting limitations of the forward model rather than the inference method. This causes a discrepancy between conventional fitting, which minimises signal residuals, and supervised training, which minimises parameter errors under the assumption that the forward model is correct. Training the GNN in a self-supervised manner, \textit{i.e.}, computing the loss between forward model predictions and measured signals on real imaging data, would more closely reproduce the outcome of conventional fitting while retaining the computational efficiency of the learning-based approach. Future work includes systematic architecture ablation studies, evaluation with additional biophysical forward models, and self-supervised training on real imaging data.

\section*{Acknowledgements}

This research was supported by the National Institute of Mental Health of the National Institutes of Health under award number 1R01MH130362-01A1 and UCL EPSRC Impact Acceleration Account to UCL 2022-26 (EP/X525649/1).

\bibliographystyle{plainnat}
\bibliography{main}

@String(CVPR= {IEEE Conf. Comput. Vis. Pattern Recog.})

@String(ECCV= {Eur. Conf. Comput. Vis.})

@String(CVPR  = {CVPR})

@String(ECCV  = {ECCV})

@book{johansen2013diffusion,
  title={Diffusion MRI: from quantitative measurement to in vivo neuroanatomy},
  author={Johansen-Berg, Heidi and Behrens, Timothy EJ},
  year={2013},
  publisher={Academic Press}
}

@article{fey2019fast,
  title={Fast graph representation learning with PyTorch Geometric},
  author={Fey, Matthias and Lenssen, Jan Eric},
  journal={arXiv preprint arXiv:1903.02428},
  year={2019}
}

@article{radhakrishnan2024practical,
  title={A practical evaluation of measures derived from compressed sensing diffusion spectrum imaging},
  author={Radhakrishnan, Hamsanandini and Zhao, Chenying and Sydnor, Valerie J and Baller, Erica B and Cook, Philip A and Fair, Damien A and Giesbrecht, Barry and Larsen, Bart and Murtha, Kristin and Roalf, David R and others},
  journal={Human Brain Mapping},
  volume={45},
  number={5},
  pages={e26580},
  year={2024},
  publisher={Wiley Online Library}
}

@article{bahdanau2014neural,
  title={Neural machine translation by jointly learning to align and translate},
  author={Bahdanau, Dzmitry and Cho, Kyunghyun and Bengio, Yoshua},
  journal={arXiv preprint arXiv:1409.0473},
  year={2014}
}

@inproceedings{landman2007diffusion,
  title={Diffusion tensor estimation by maximizing Rician likelihood},
  author={Landman, Bennett and Bazin, Pierre-Louis and Prince, Jerry},
  booktitle={2007 IEEE 11th International Conference on Computer Vision},
  pages={1--8},
  year={2007},
  organization={IEEE}
}

@article{stejskal1965spin,
  title={Spin diffusion measurements: spin echoes in the presence of a time-dependent field gradient},
  author={Stejskal, Edward O and Tanner, John E},
  journal={The journal of chemical physics},
  volume={42},
  number={1},
  pages={288--292},
  year={1965},
  publisher={American Institute of Physics}
}

@article{sotiropoulos2013advances,
  title={Advances in diffusion MRI acquisition and processing in the Human Connectome Project},
  author={Sotiropoulos, Stamatios N and Jbabdi, Saad and Xu, Junqian and Andersson, Jesper L and Moeller, Steen and Auerbach, Edward J and Glasser, Matthew F and Hernandez, Moises and Sapiro, Guillermo and Jenkinson, Mark and others},
  journal={Neuroimage},
  volume={80},
  pages={125--143},
  year={2013},
  publisher={Elsevier}
}

@article{alfaro2018image,
  title={Image processing and Quality Control for the first 10,000 brain imaging datasets from UK Biobank},
  author={Alfaro-Almagro, Fidel and Jenkinson, Mark and Bangerter, Neal K and Andersson, Jesper LR and Griffanti, Ludovica and Douaud, Gwena{\"e}lle and Sotiropoulos, Stamatios N and Jbabdi, Saad and Hernandez-Fernandez, Moises and Vallee, Emmanuel and others},
  journal={Neuroimage},
  volume={166},
  pages={400--424},
  year={2018},
  publisher={Elsevier}
}

@article{elaldi2024equivariant,
  title={Equivariant spatio-hemispherical networks for diffusion MRI deconvolution},
  author={Elaldi, Axel and Gerig, Guido and Dey, Neel},
  journal={Advances in Neural Information Processing Systems},
  volume={37},
  pages={52095--52126},
  year={2024}
}

@article{bronstein2021geometric,
  title={Geometric deep learning: Grids, groups, graphs, geodesics, and gauges},
  author={Bronstein, Michael M and Bruna, Joan and Cohen, Taco and Veli{\v{c}}kovi{\'c}, Petar},
  journal={arXiv preprint arXiv:2104.13478},
  year={2021}
}

@article{karimi2024diffusion,
  title={Diffusion MRI with machine learning},
  author={Karimi, Davood and Warfield, Simon K},
  journal={Imaging Neuroscience},
  volume={2},
  pages={1--55},
  year={2024},
  publisher={MIT Press}
}

@inproceedings{satorras2021n,
  title={E (n) equivariant graph neural networks},
  author={Satorras, V{\i}ctor Garcia and Hoogeboom, Emiel and Welling, Max},
  booktitle={International conference on machine learning},
  pages={9323--9332},
  year={2021},
  organization={PMLR}
}

@inproceedings{gilmer2017neural,
  title={Neural message passing for quantum chemistry},
  author={Gilmer, Justin and Schoenholz, Samuel S and Riley, Patrick F and Vinyals, Oriol and Dahl, George E},
  booktitle={International conference on machine learning},
  pages={1263--1272},
  year={2017},
  organization={Pmlr}
}

@article{scarselli2008graph,
  title={The graph neural network model},
  author={Scarselli, Franco and Gori, Marco and Tsoi, Ah Chung and Hagenbuchner, Markus and Monfardini, Gabriele},
  journal={IEEE transactions on neural networks},
  volume={20},
  number={1},
  pages={61--80},
  year={2008},
  publisher={IEEE}
}

@inproceedings{chen2020estimating,
  title={Estimating tissue microstructure with undersampled diffusion data via graph convolutional neural networks},
  author={Chen, Geng and Hong, Yoonmi and Zhang, Yongqin and Kim, Jaeil and Huynh, Khoi Minh and Ma, Jiquan and Lin, Weili and Shen, Dinggang and Yap, Pew-Thian and UNC/UMN Baby Connectome Project Consortium},
  booktitle={International Conference on Medical Image Computing and Computer-Assisted Intervention},
  pages={280--290},
  year={2020},
  organization={Springer}
}

@inproceedings{chen2022hybrid,
  title={Hybrid graph transformer for tissue microstructure estimation with undersampled diffusion MRI data},
  author={Chen, Geng and Jiang, Haotian and Liu, Jiannan and Ma, Jiquan and Cui, Hui and Xia, Yong and Yap, Pew-Thian},
  booktitle={International Conference on Medical Image Computing and Computer-Assisted Intervention},
  pages={113--122},
  year={2022},
  organization={Springer}
}

@inproceedings{esteves2018learning,
  title={Learning so (3) equivariant representations with spherical cnns},
  author={Esteves, Carlos and Allen-Blanchette, Christine and Makadia, Ameesh and Daniilidis, Kostas},
  booktitle={Proceedings of the european conference on computer vision (ECCV)},
  pages={52--68},
  year={2018}
}

@article{cohen2018spherical,
  title={Spherical cnns},
  author={Cohen, Taco S and Geiger, Mario and K{\"o}hler, Jonas and Welling, Max},
  journal={arXiv preprint arXiv:1801.10130},
  year={2018}
}

@article{kerkela2024spherical,
  title={Spherical convolutional neural networks can improve brain microstructure estimation from diffusion MRI data},
  author={Kerkel{\"a}, Leevi and Seunarine, Kiran and Szczepankiewicz, Filip and Clark, Chris A},
  journal={Frontiers in Neuroimaging},
  volume={3},
  pages={1349415},
  year={2024},
  publisher={Frontiers Media SA}
}

@inproceedings{goodwin2022can,
  title={How can spherical CNNs benefit ML-based diffusion MRI parameter estimation?},
  author={Goodwin-Allcock, Tobias and McEwen, Jason and Gray, Robert and Nachev, Parashkev and Zhang, Hui},
  booktitle={International Workshop on Computational Diffusion MRI},
  pages={101--112},
  year={2022},
  organization={Springer}
}

@inproceedings{sedlar2021spherical,
  title={A spherical convolutional neural network for white matter structure imaging via dMRI},
  author={Sedlar, Sara and Alimi, Abib and Papadopoulo, Th{\'e}odore and Deriche, Rachid and Deslauriers-Gauthier, Samuel},
  booktitle={International Conference on Medical Image Computing and Computer-Assisted Intervention},
  pages={529--539},
  year={2021},
  organization={Springer}
}

@inproceedings{sedlar2021diffusion,
  title={Diffusion MRI fiber orientation distribution function estimation using voxel-wise spherical U-net},
  author={Sedlar, Sara and Papadopoulo, Th{\'e}odore and Deriche, Rachid and Deslauriers-Gauthier, Samuel},
  booktitle={Computational Diffusion MRI: International MICCAI Workshop, Lima, Peru, October 2020},
  pages={95--106},
  year={2021},
  organization={Springer}
}

@article{rot2025real,
  title={Real-time, inline quantitative MRI enabled by scanner-integrated machine learning: a proof of principle with NODDI},
  author={Rot, Samuel and Dragonu, Iulius and Triantafyllou, Christina and Grech-Sollars, Matthew and Papadaki, Anastasia and Mancini, Laura and Wastling, Stephen and Steeden, Jennifer and Thornton, John and Yousry, Tarek and others},
  journal={arXiv preprint arXiv:2507.12632},
  year={2025}
}

@inproceedings{legouhy2024eddeep,
  title={Eddeep: Fast eddy-current distortion correction for diffusion MRI with deep learning},
  author={Legouhy, Antoine and Callaghan, Ross and Stee, Whitney and Peigneux, Philippe and Azadbakht, Hojjat and Zhang, Hui},
  booktitle={International Conference on Medical Image Computing and Computer-Assisted Intervention},
  pages={152--161},
  year={2024},
  organization={Springer}
}

@article{szafer1995theoretical,
  title={Theoretical model for water diffusion in tissues},
  author={Szafer, Aaron and Zhong, Jianhui and Gore, John C},
  journal={Magnetic resonance in medicine},
  volume={33},
  number={5},
  pages={697--712},
  year={1995},
  publisher={Wiley Online Library}
}

@article{novikov2019quantifying,
  title={Quantifying brain microstructure with diffusion MRI: Theory and parameter estimation},
  author={Novikov, Dmitry S and Fieremans, Els and Jespersen, Sune N and Kiselev, Valerij G},
  journal={NMR in Biomedicine},
  volume={32},
  number={4},
  pages={e3998},
  year={2019},
  publisher={Wiley Online Library}
}

@article{zong2024attention,
  title={Attention-Based Q-Space Deep Learning Generalized for Accelerated Diffusion Magnetic Resonance Imaging},
  author={Zong, Fangrong and Zhu, Zaimin and Zhang, Jiayi and Deng, Xiaofeng and Li, Zhuangzhuang and Ye, Chuyang and Liu, Yong},
  journal={IEEE Journal of Biomedical and Health Informatics},
  year={2024},
  publisher={IEEE}
}

@article{ewert2024geometric,
  title={Geometric deep learning for diffusion mri signal reconstruction with continuous samplings (discus)},
  author={Ewert, Christian and K{\"u}gler, David and Stirnberg, R{\"u}diger and Koch, Alexandra and Yendiki, Anastasia and Reuter, Martin},
  journal={Imaging neuroscience},
  volume={2},
  pages={1--18},
  year={2024},
  publisher={MIT Press}
}

@article{basser1994mr,
  title={MR diffusion tensor spectroscopy and imaging},
  author={Basser, Peter J and Mattiello, James and LeBihan, Denis},
  journal={Biophysical journal},
  volume={66},
  number={1},
  pages={259--267},
  year={1994},
  publisher={Elsevier}
}

@article{jensen2005diffusional,
  title={Diffusional kurtosis imaging: the quantification of non-gaussian water diffusion by means of magnetic resonance imaging},
  author={Jensen, Jens H and Helpern, Joseph A and Ramani, Anita and Lu, Hanzhang and Kaczynski, Kyle},
  journal={Magnetic Resonance in Medicine: An Official Journal of the International Society for Magnetic Resonance in Medicine},
  volume={53},
  number={6},
  pages={1432--1440},
  year={2005},
  publisher={Wiley Online Library}
}

@article{westin2016q,
  title={Q-space trajectory imaging for multidimensional diffusion MRI of the human brain},
  author={Westin, Carl-Fredrik and Knutsson, Hans and Pasternak, Ofer and Szczepankiewicz, Filip and {\"O}zarslan, Evren and van Westen, Danielle and Mattisson, Cecilia and Bogren, Mats and O'donnell, Lauren J and Kubicki, Marek and others},
  journal={Neuroimage},
  volume={135},
  pages={345--362},
  year={2016},
  publisher={Elsevier}
}

@article{zhang2012noddi,
  title={{NODDI}: practical in vivo neurite orientation dispersion and density imaging of the human brain},
  author={Zhang, Hui and Schneider, Torben and Wheeler-Kingshott, Claudia A.~M.~Gandini and Alexander, Daniel C.},
  journal={NeuroImage},
  volume={61},
  number={4},
  pages={1000--1016},
  year={2012}
}

@article{golkov2016q,
  title={Q-space deep learning: twelve-fold shorter and model-free diffusion MRI scans},
  author={Golkov, Vladimir and Dosovitskiy, Alexey and Sperl, Jonathan I and Menzel, Marion I and Czisch, Michael and S{\"a}mann, Philipp and Brox, Thomas and Cremers, Daniel},
  journal={IEEE transactions on medical imaging},
  volume={35},
  number={5},
  pages={1344--1351},
  year={2016},
  publisher={IEEE}
}

@article{park2021diffnet,
  title={DIFFnet: diffusion parameter mapping network generalized for input diffusion gradient schemes and b-value},
  author={Park, Juhyung and Jung, Woojin and Choi, Eun-Jung and Oh, Se-Hong and Jang, Jinhee and Shin, Dongmyung and An, Hongjun and Lee, Jongho},
  journal={IEEE Transactions on Medical Imaging},
  volume={41},
  number={2},
  pages={491--499},
  year={2021},
  publisher={IEEE}
}

@inproceedings{qi2017pointnet,
  title={{PointNet}: Deep Learning on Point Sets for 3D Classification and Segmentation},
  author={Qi, Charles R. and Su, Hao and Mo, Kaichun and Guibas, Leonidas J.},
  booktitle={Proceedings of the IEEE Conference on Computer Vision and Pattern Recognition (CVPR)},
  pages={652--660},
  year={2017}
}

@article{zaheer2017deep,
  title={Deep sets},
  author={Zaheer, Manzil and Kottur, Satwik and Ravanbakhsh, Siamak and Poczos, Barnabas and Salakhutdinov, Russ R and Smola, Alexander J},
  journal={Advances in neural information processing systems},
  volume={30},
  year={2017}
}

\end{document}